\documentclass[aps,prl,twocolumn,superscriptaddress,showpacs,amsmath,amssymb]{revtex4}

\usepackage{graphicx}
\usepackage{dcolumn}
\usepackage{bm}
\def\prn#1{{\left(#1\right)}}

\def\sbrk#1{{\left[#1\right]}}

\def\clebsch#1#2#3#4#5#6{\langle #1 \, #2 \, #3 \, #4 \vert #5 \, #6 \rangle}

\def\sbrk#1{{\left[#1\right]}}

\def\prn#1{{\left(#1\right)}}

\newcommand{\TPOne}{${\rm 6s6p \:} ^3{\rm P}_1\:$}
\newcommand{\TPZero}{${\rm 6s6p \:} ^3{\rm P}_0\:$}
\newcommand{\TDOne}{${\rm 5d6s \:} ^3{\rm D}_1\:$}
\newcommand{\SPOne}{${\rm 6s6p \:} ^1{\rm P}_1\:$}
\newcommand{\SSZeroToTDOne}{\mbox{${\rm 6s^2}~ ^1{\rm S}_0~\rightarrow ~{\rm 5d6s
}~ ^3{\rm D}_1\:$}}

\def\clebsch#1#2#3#4#5#6{\langle #1 \, #2 \, #3 \, #4 \vert #5 \, #6 \rangle}

\begin{document}


\title{Observation of a Large Atomic Parity Violation  Effect in Ytterbium}

\author{K. Tsigutkin}
\email{tsigutkin@berkeley.edu}%
\author{D. Dounas-Frazer}%
\author{A. Family}%
\author{J. E. Stalnaker}%
\altaffiliation[Present address: ]{Department of Physics and
Astronomy, Oberlin College, Oberlin, OH 44074}%
\affiliation{Department of Physics, University of California at Berkeley,\\
Berkeley, CA 94720-7300}%
\author{V. V. Yashchuk}%
\affiliation{Advanced Light Source Division, Lawrence Berkeley
National Laboratory, Berkeley CA 94720}
\author{D. Budker}%
\affiliation{Department of Physics, University of California at Berkeley,\\
Berkeley, CA 94720-7300}
\affiliation{Nuclear Science Division, Lawrence Berkeley National Laboratory, Berkeley, California 94720}%

\date{\today}

\begin{abstract}
Atomic parity violation has been observed in the \SSZeroToTDOne
408-nm forbidden transition of ytterbium. The parity-violating
amplitude is found to be two orders of magnitude larger than in
cesium, where the most precise experiments to date have been
performed. This is in accordance with theoretical predictions and
constitutes the largest atomic parity-violating amplitude yet
observed. This also opens the way to future measurements of
neutron skins and anapole moments by comparing parity-violating
amplitudes for various isotopes and hyperfine components of the
transition.
\end{abstract}

\pacs{11.30.Er, 32.80.Ys}
\keywords{Suggested keywords}
\maketitle

Atomic Parity Violation (APV) experiments are a powerful tool in
the study of electroweak interactions (see, for example, review
\cite{Guena2005}). The electroweak parameter of utmost importance
in APV experiments is the weak charge $Q_W$, associated with the
exchange of the $Z_0$ boson between an atomic electron and the
nucleus. The most accurate APV experiments were performed using Cs
atomic beam and yielded a value of the $Q_W$ of Cs having an
experimental and theoretical uncertainties of 0.35\% \cite{Cs97}
and 0.20\% \cite{Cs2009th}, respectively, providing a stringent
test of the Standard Model (SM) at low momentum transfer
($\approx$~MeV/c). However, it has not yet been possible to test
an important prediction of the SM concerning the variation of
$Q_W$ along a chain of isotopes. It has been suggested
\cite{Dzuba86} that rare-earth atoms may be good candidates for
APV experiments because they have chains of stable isotopes, and
the APV effects may be enhanced due to the proximity of
opposite-parity levels. While the accuracy of atomic calculations
is unlikely to ever approach that achieved for atoms with a single
valence electron, ratios of PV-amplitudes between different
isotopes should provide ratios of weak charges, without involving,
to first approximation, any atomic-structure calculations.

The present experiment is inspired by the prediction
\cite{demille95} supported by further theoretical work of
\cite{Por95,Das97}, that the PV-amplitude in the chosen transition
is $\approx$100 times larger than that in Cs. The motivation for
PV-experiments in Yb
is probing low-energy nuclear physics by comparing PV-effects on
either a chain of naturally occurring Yb isotopes, or in different
hyperfine components for the same odd-neutron-number isotope. The
ratio of PV amplitudes for two isotopes of the same element is
sensitive to the \emph{neutron distributions} within the nucleus.
The difference between PV amplitudes measured on two different
hyperfine lines belonging to the same transition is a
manifestation of nuclear-spin-dependent APV, which is sensitive to
the nuclear \emph{anapole moments} (see, for example, reviews
\cite{Gin2004,Hax2001}) that arise from weak interactions between
the nucleons. As the precision of the experiment increases, a
sensitive test of the Standard Model may also become possible
\cite{nSkin}.

Here we report on experimental verification of the predicted
PV-amplitude enhancement in Yb using a measurement of the APV
amplitude for $^{174}$Yb.

\begin{figure}[!htb]
\resizebox{0.35\textwidth}{!}{%
  \includegraphics{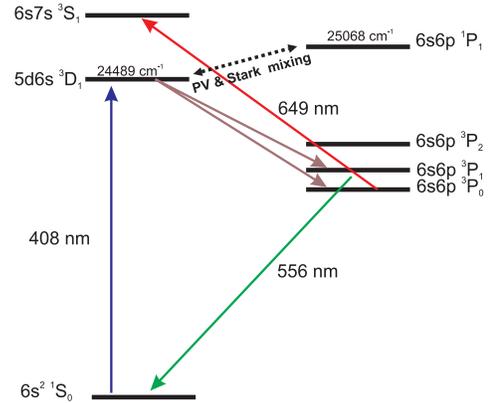}}
\caption{(color online) Low-lying energy eigenstates of Yb and
transitions relevant to the APV experiment.} \label{levels}
\end{figure}

The idea of the experiment is to excite the forbidden 408-nm
transition (Fig. \ref{levels}) with resonant laser light in the
presence of a quasi-static electric field. The PV-amplitude of
this transition arises due to PV-mixing of the \TDOne and \SPOne
states. The purpose of the electric field is to provide a
reference transition amplitude due to Stark-mixing of the same
states, interfering with the PV amplitude. In such interference
method \cite{Bou75,Con79}, one is measuring the part of the
transition probability that is linear in both the reference
Stark-induced amplitude and the PV amplitude. In addition to
enhancing the PV-dependent signal, employing the Stark-PV
interference technique provides for all-important reversals
allowing one to separate the PV effects from various systematics.
\begin{figure}[!htb]
\resizebox{0.47\textwidth}{!}{%
  \includegraphics{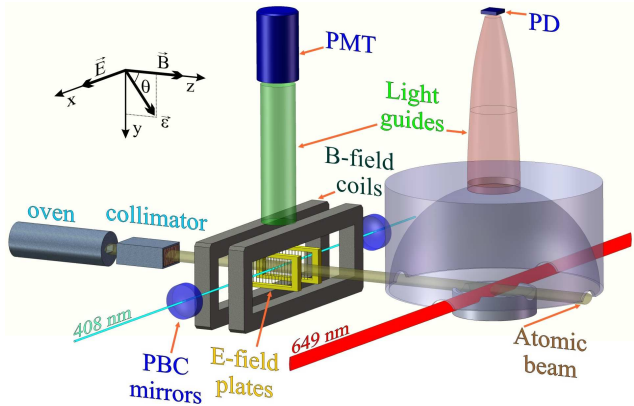}
} \caption{(color online) Orientation of fields for PV-Stark
interference experiment and schematic of the present APV
apparatus. Not shown is the vacuum chamber containing all the
depicted elements, except the photomultiplier (PMT) and the
photodiode (PD). PBC--power buildup cavity. Light is applied
collinearly with \textbf{x}.} \label{apparatus}
\end{figure}

The configuration which is used for the Stark-interference
experiment is shown in Fig. \ref{apparatus}. The electric field,
\textbf{E} is applied collinearly with the propagation axis
(\textbf{x}) of the linearly-polarized resonant light beam, while
the magnetic field, \textbf{B} is directed along \textbf{z}. The
pseudo-scalar quantity which manifests PV is:
\begin{equation}
\prn{\bf{\mathcal{E}} \cdot
\bf{B}}\prn{\sbrk{\bf{E}\times\mathcal{E}}\cdot
\bf{B}},\label{Invariant}
\end{equation}
where $\bf{\mathcal{E}}$ is the electric field of the light. The
APV effect vanishes when the angle $\theta$ between the light
polarization and the magnetic field approaches a value which is a
multiple of $\pi/2$.

This field arrangement is such that the M1 transition amplitude
and Stark-induced amplitudes are out of phase \cite{Dre85}. Thus,
the M1-Stark interference is suppressed. Additional suppression is
provided by the use of a power-build-up cavity. The M1 transition
amplitude proportional to $\bf{k}\times \bf{\mathcal{E}}$ vanishes
to the degree that the field in the cavity is a standing wave, and
the net wavevector $\bf{k}$ is suppressed.

For an isotope with zero nuclear spin $I$, there are three
Zeeman-split components of the transition. A Stark-induced
transition amplitude is generally expressed in terms of real
scalar ($\alpha$), vector ($\beta$), and tensor ($\gamma$)
transition polarizabilities \cite{Bou75,Bow99}, however, for the
case of a $J=0\rightarrow J'=1$ transition, only the vector
transition polarizability contributes. Assuming that the magnetic
field is strong enough to resolve the Zeeman components of the
transition and selecting the quantization axis along the magnetic
field, we obtain the following transition rates:
\begin{eqnarray}  \label{even_rates}
\mathcal{R}_{\Delta M=0}=\frac{8\pi}{c}\mathcal{I}\sbrk{\beta^2
E^2\sin^2\theta+2\zeta\,\beta E  \sin \theta \cos \theta},
\\\mathcal{R}_{\Delta M=\pm 1}=\frac{4\pi}{c}\mathcal{I}\sbrk{\beta^2
E^2\cos^2\theta-2\zeta\,\beta E  \sin \theta \cos
\theta},\label{even_rates2}
\end{eqnarray}
where $\mathcal{I}$ is the light intensity. Here $\zeta$
characterizes the PV-induced electric-dipole transition amplitude
between states with total angular momenta and projections $F,M$
and $F',M'$:
\begin{eqnarray}
A^{PV}_{FMF'M'}=i\zeta_{FF'}(-1)^{q}\mathcal{E}_{q}\langle
F,M,1,-q|F',M'\rangle~,
\end{eqnarray}
where $q=M-M'$ labels the spherical component and
$\clebsch{F,}{M,}{1,}{-q}{F',}{M'}$ is a Clebsch-Gordan
coefficient. In expressions (\ref{even_rates},\ref{even_rates2}),
we neglect the term quadratic in PV mixing. Using the theoretical
value of $\zeta\simeq 10^{-9}~\text{ea}_0$ \cite{Por95} combined
with the measured $|\beta |= 2.24_{-0.12}^{+0.07}\times
10^{-8}~\text{ea}_0$/(V/cm) \cite{Bow99,Sta2006}, the expected
relative strength of the PV-effect, $2\zeta/\beta E$, is $\sim
10^{-4}$, for $\theta=\pi/4$ and $E=1$~kV/cm.

The transition rates (\ref{even_rates},\ref{even_rates2}) are
detected by measuring the population of the \TPZero state, where
65\% of the atoms excited to the \TDOne state decay spontaneously
(Fig. \ref{levels}). This is done by resonantly exciting the atoms
with 649-nm light to the ${\rm 6s7s} \:^3{\rm S}_1$ state
downstream from the main interaction region and collecting the
fluorescence resulting from the decay of this state back to
\TPZero state, and also to \TPOne and ${\rm 6s6p \:} ^3{\rm
P}_2\:$ states. As long as the 408-nm transition is not saturated,
the fluorescence intensity measured in the probe region is
proportional to the rate of that transition.

\begin{figure}[!htb]
\resizebox{0.47\textwidth}{!}{%
  \includegraphics{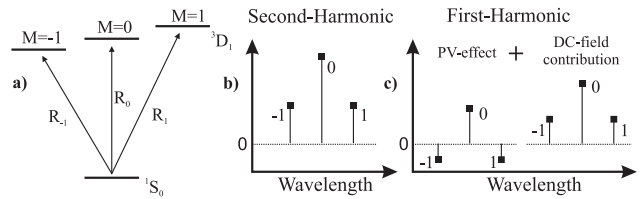}
} \caption{Discrimination of the PV-effect by E-field modulation
under static magnetic field.} \label{lshape}
\end{figure}

In order to isolate the Stark-PV interference term in the
transition rate from the dominant Stark-induced transition rate,
we harmonically modulate the applied electric field. The dominant
Stark-induced rate has a static component and a component
oscillating at twice the modulation frequency, while the Stark-PV
interference term oscillates at the first harmonic. The frequency
discrimination is performed using lock-in amplifiers. For an
arbitrary angle of the light polarization $\theta$, there are
generally three Zeeman components of the transition present while
scanning over the profile as shown in Fig. \ref{lshape}a. The
first-harmonic signal due to Stark-PV interference has a
characteristic signature: the sign of the phase of the modulation
of the two extreme components of the transition is opposite to
that of the central component. The second-harmonic signal provides
a reference for the lineshape since it is free from interference
effects linear in \textbf{E} (Fig. \ref{lshape}b). If, in addition
to the oscillating electric field, there is also a DC component
present in the applied field, a signature identical to that in the
second harmonic will also appear in the first harmonic, Fig.
\ref{lshape}c. The latter can be used to increase the
first-harmonic signal above the noise, which makes the profile
analysis more reliable.

A schematic of the Yb-APV apparatus is shown in Fig.
\ref{apparatus}. A beam of Yb atoms is produced (inside of a
vacuum chamber with a residual pressure of $\approx 3 \times
10^{-6}~{\rm Torr}$) with an effusive source, which is a
stainless-steel oven loaded with Yb metal, operating at
$500^{\circ}{\rm C}$. The oven is outfitted with a multi-slit
nozzle, and there is an external vane collimator reducing the
spread of the atomic beam in the horizontal direction. The
resulting Doppler width of the 408-nm transition is $\approx 12 \,
{\rm MHz}$ \cite{Sta2006}.

Downstream from the collimator, the atoms enter the main
interaction region where the Stark- and PV-induced transition
takes place. Up to 80~mW of light at the transition wavelength of
$408.345 \, {\rm nm}$  in vacuum is produced by frequency doubling
the output of a Ti:Sapphire laser (Coherent $899$).

The 408-nm light is coupled into a power buildup cavity (PBC)
inside the vacuum chamber. The finesse and the circulating power
of the PBC are measured to be up to $\mathcal{F}=9000$ and
$P=8$~W, respectively. The laser is locked to the PBC using the
FM-sideband technique \cite{Drev83}. In order to remove frequency
excursions of the PBC in the acoustic range, the cavity is locked
to a more stable confocal Fabry-P\'{e}rot \'{e}talon, once again
using the FM-sideband technique.  This stable scannable cavity
provides the master frequency, with the power-build-up cavity
serving as the secondary master for the laser.

The magnetic field is generated by a pair of rectangular coils
designed to produce a uniform magnetic field up to $100\ $G: 1\%
non-uniformity over the volume with the dimensions of $1\times
1\times 1~\text{cm}^3$ in the interaction region. Additional coils
placed outside of the vacuum chamber compensate for the external
magnetic fields down to $10~\text{mG}$ at the interaction region.
The residual B-field of this magnitude does not have an impact on
the PV-effect measurements, since its contribution is measured
using the field reversals (see below).

The electric field is generated with two wire-frame electrodes
separated by 2 cm. The copper electrode frames support arrays of
0.2-mm dia. gold-plated wires. This design allows to reduce the
stray charges accumulated on the electrodes by minimizing the
surface area facing the atomic beam, thus minimizing stray
electric fields. AC voltage up to 10~kV at a frequency of 76.2~Hz
is supplied to the E-field electrodes by a home-built high-voltage
amplifier. An additional DC bias voltage up to 100~V can be added.

Light emitted from the interaction region at 556$\ $nm is
collected with a light guide and detected with a photomultiplier
tube. This signal is used for initial selection of the atomic
resonance and for monitoring purposes. Fluorescent light from the
probe region is collected onto a light guide using two optically
polished curved aluminum reflectors and registered with a cooled
photodetector~(PD). The PD consists of a large-area ($1\times
1~\text{cm}^2$) Hamamatsu photodiode connected to a 1-G$\Omega$
transimpedance pre-amplifier, both contained in a cooled housing
(temperatures down to $-15^{\circ}{\rm C}$). The pre-amp's
bandwidth is 1~kHz and the output noise is $\sim 1$~mV (rms). The
649-nm excitation light is derived from a single-frequency diode
laser (New Focus Vortex) producing $\approx 1.2\ $mW of cw output,
high enough to saturate the {\mbox{${\rm 6s6p \:} ^3{\rm
P}_0~\rightarrow ~{\rm 6s7s} \:^3{\rm S}_1$} transition. A drift
of the laser frequency is eliminated by locking the diode laser to
a transfer cavity, in turn locked to a frequency-stabilized He-Ne
laser.

The signals from the PMT and PD are fed into lock-in amplifiers
for frequency discrimination and averaging. The typical time of a
single spectral profile acquisition is 20~s. The first- and
second-harmonic signals are registered simultaneously, which
reduces the influence of slow drifts, such as instabilities of the
atomic-beam flux. The modulation frequency is limited by several
factors. Thermal distribution of atomic velocities in the beam
causes a spread  (of $\approx$~2~ms) in the time of flight between
the interaction region and the probe region. This, along with the
finite bandwidth of the PD, leads to a reduction of the
signal-modulation contrast. The choice of the modulation frequency
of 76.2~Hz is a tradeoff between this contrast degradation and the
requirement of lock-in detection over many modulation periods.

\begin{figure}[!htb]
\resizebox{0.5\textwidth}{!}{%
  \includegraphics[bb=313 210 730 390]{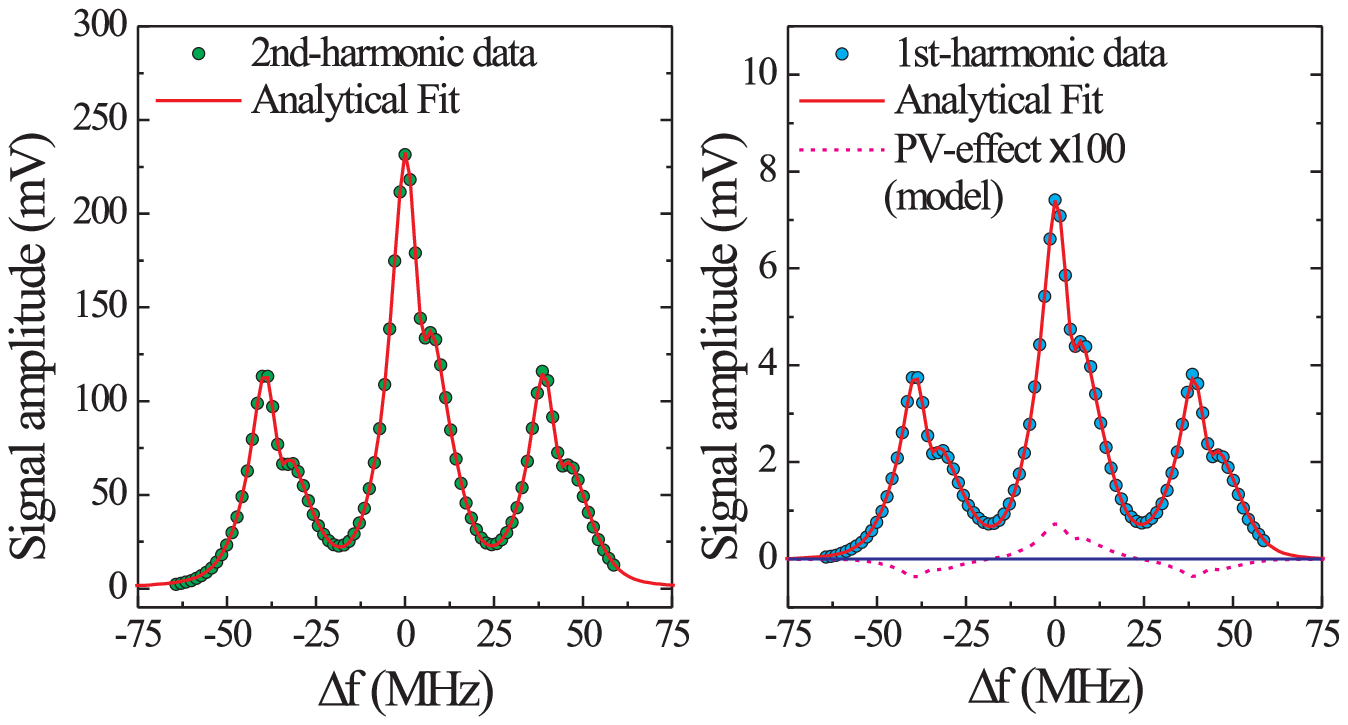}
} \caption{(color online) A profile of the B-field-split 408-nm
spectral line of $^{174}$Yb recorded at 1st- and 2nd-harmonic of
the modulation. Also a simulated PV-contribution is shown for
clarity. $\tilde{E}$=5~kV/cm; DC offset=40~V/cm; $\theta=\pi/4$;
an effective integration time is 10~s per point.}
\label{Exp_lshape}
\end{figure}

In Fig.~\ref{Exp_lshape} a profile of the B-field-split 408-nm
spectral line of the $^{174}$Yb isotope is shown. The 649-nm
light-induced fluorescence was recorded during a typical
integration run. Statistical error bars are smaller than the size
of the points in the figure. The peculiar asymmetric line shape of
the Zeeman components is a result of the dynamic Stark effect
\cite{Sta2006}. The profiles are fit to an approximated analytic
function. The fit amplitudes of the three peaks yield the
fluorescence amplitudes for the different Zeeman components at
1-st and 2-nd harmonics of the modulation. Ideally, in the absence
of apparatus imperfections and systematic effects, the following
combination of amplitude ratios between the 1-st and 2-nd
harmonics yields the PV-interference effect:
\begin{eqnarray}  \label{ratios}
\mathcal{K}=\frac{\mathcal{A}_{-1}^{1st}}{\mathcal{A}_{-1}^{2nd}}+
\frac{\mathcal{A}_{+1}^{1st}}{\mathcal{A}_{+1}^{2nd}}-2\frac{\mathcal{A}_{0}^{1st}}{\mathcal{A}_{0}^{2nd}}
=\frac{16\zeta}{\beta\tilde{E}},
\end{eqnarray}
where $\mathcal{A}_{\pm 1}$, $\mathcal{A}_{0}$ are the amplitudes
of the respective Zeeman components and $\tilde{E}$ is the
amplitude of modulating electric field.

The detailed analysis of an impact of the apparatus imperfections
and systematic effects on the accuracy of the measurements will be
presented elsewhere. Here we address briefly the principles of
this analysis. The PV-effect is discriminated from other effects
by recording the spectral profiles for different combinations of
the $B$-field and the light polarization angle $\theta$, and by
isolating the part of the measured values of $\mathcal{K}$ that
has a correct PV-response upon the reversals. In addition we
artificially impose exaggerated combinations of imperfections and
measure their effect on $\mathcal{K}$. Then, by scaling down these
contributions we estimate the residual uncertainties in the
PV-measurements. Such experiments showed a negligibly small
contribution of the imperfections compared to the present accuracy
of the PV-effect determination (see below).

\begin{figure}[!htb]
\resizebox{0.5\textwidth}{!}{%
  \includegraphics[bb=260 270 520 395]{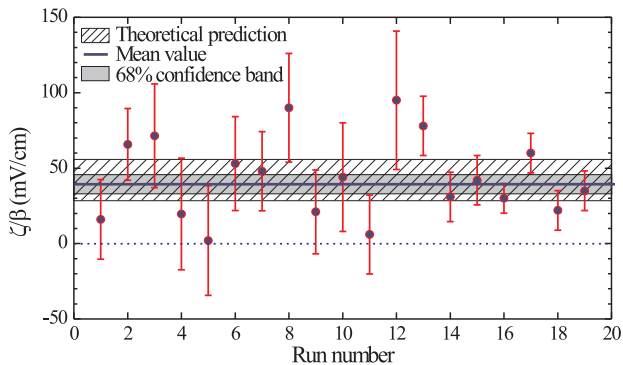}
} \caption{(color online) The PV-interference parameter $\zeta
/\beta$. Mean value:
$39(4)_{\text{stat.}}(5)_{\text{syst.}}~\text{mV/cm}$, $|\zeta
|=8.7\pm 1.4\times 10^{-10}$~ea$_0$.} \label{APV}
\end{figure}

In Fig.~\ref{APV}, the PV-interference parameter $\zeta /\beta$ is
shown as determined in 19 consecutive runs ($\sim$60 hours of
integration). Its mean value is
$39(4)_{\text{stat.}}(5)_{\text{syst.}}~\text{mV/cm}$, which is in
agreement with the theoretical predictions. Thus, $|\zeta |=8.7\pm
1.4\times 10^{-10}$~ea$_0$, which is the largest APV amplitude
observed so far. This confirms the predicted enhancement of the
PV-effect in Yb.

The present measurement accuracy is not yet sufficient to observe
the isotopic and hyperfine differences in the PV-amplitude. It
must be better than $\approx 1\%$ for PV-amplitude in a single
transition \cite{nSkin,AnapoleKozlov,Das99}. We found that the
main factors limiting the present accuracy are fluctuations of the
electric field in the interaction region (due to stray fields and
HV-amplifier noises), and frequency excursions of the
Fabry-P\'{e}rot \'{e}talon serving as a frequency reference for
the optical system. A direct impact of these factors on the
spectral profiles has been observed, thus, leading to errors not
only in the APV measurements, but also in the study of systematic
effects and apparatus imperfections. This accounts for the
relatively large systematic uncertainty of the PV-parameter. In
the course of the APV measurements, several improvements have been
implemented targeting these noise sources. They have demonstrated
a possibility to reduce the measurement errors substantially. This
is seen in Fig.~\ref{APV}, where the last six measurements exhibit
higher accuracy than the rest. An upgrade of the apparatus is
underway aimed at eliminating the noise sources, which will open
ways to the measurements of neutron skins and anapole moments.

The authors acknowledge helpful discussions with and important
contributions of M. A. Bouchiat, C. J. Bowers, E. D. Commins, B.
P. Das, D. DeMille, A. Dilip, S. J. Freedman, J. S. Guzman, G.
Gwinner, M. G. Kozlov, S. M. Rochester, and M. Zolotorev. This
work has been supported by NSF.

\bibliography{Yb_APV}

\end{document}